# Continuous-wave 1.55 µm diode-pumped surface emitting semiconductor laser for broadband multiplex spectroscopy

**M. Jacquemet, N. Picqué, G. Guelachvili,**

*Laboratoire de Photophysique Moléculaire, CNRS ; Univ Paris-Sud, Bât. 350, F-91405 Orsay Cedex, France*

**A. Garnache**

*CEM2, CNRS UMR5507, Université Montpellier 2, Montpellier, France*

**I. Sagnes, M. Strassner, C. Symonds**

*Laboratoire de Photonique et de Nanostructures, LPN-CNRS, Route de Nozay, 91460 Marcoussis, France*

**Optics Letters, in press, 2007**

Corresponding author:
Dr. Nathalie Picqué,
Laboratoire de Photophysique Moléculaire
Unité Propre du CNRS, Université de Paris-Sud, Bâtiment 350
91405 Orsay Cedex, France
Website: http://www.laser-fts.org
Phone nb: 33 1 69 15 66 49
Fax nb: 33 1 69 15 75 30
Email: nathalie.picque@ppm.u-psud.fr

Abstract: A room temperature operating Vertical External Cavity Surface Emitting Laser is applied around 1550 nm to intracavity laser absorption spectroscopy analyzed by time-resolved Fourier transform interferometry. At an equivalent pathlength of 15 km, the high resolution spectrum of the semiconductor disk laser emission covers 17 nm simultaneously. A noise equivalent absorption coefficient at one second averaging equal to $1.5 \ 10^{-10}$ cm$^{-1}$.Hz$^{-1/2}$ per spectral element is reported for the 65 km longest path length employed.

Over the past few years, spectroscopy in the near-infrared region of the spectrum has benefited from the substantial development of laser technology associated with the 1.3 and 1.5 µm telecommunication bands. Compact, stable, and high-power sources have been developed for various commercial and technological applications and have thus served as an impetus for active research in this wavelength region. Seat of the weak overtone molecular transitions, the 1.5 µm region has prompted the need for high sensitivity spectroscopic techniques, reaching now the ppb detection levels. One of the prevailing quests presently aims at combining these sensitivities with rapid diagnostic times and/or multi-species gas analysis.

Since its introduction in the 1970's, Intracavity Laser Absorption Spectroscopy (ICLAS) has been recognized [1] as one of the most sensitive techniques. The absorbing sample is placed inside the resonator of a multimode laser, the gain spectrum of which is broader than the absorbing transitions to be detected. From the beginning of the laser pulse to the time of observation $t_g$ (generation time), absorption follows the Lambert-Beer law with an equivalent path length equal to $(\ell_{cell}/L_{Res}) \, c \, t_g$ where $c$ is the velocity of light, $\ell_{cell}$ the length of the cell containing the absorbing sample and $L_{Res}$ the resonator length. The main advantages of ICLAS are sensitivity due to kilometric absorption path lengths and large spectral detection bandwidth due to the broad spectrum of multimode lasers. ICLAS experiments have been performed with various types of broadband laser sources. In particular, Optically-Pumped Vertical-External-Cavity Surface-Emitting Lasers (OP-VECSELs) on GaAs substrates have proved their efficiency in the 1 µm region [2,3] and recently around 2.3 µm with Sb-based materials [4], pumped by commercial low power GaAs-based diode lasers.

Indeed, VECSELs constitute a promising family of versatile solid-state lasers for spectroscopic purposes, with their low laser threshold, circular diffraction-limited output beam [5] and large tunability and gain bandwidth [2-4]. A recent review is included in [6]. In the 1.55 µm water-free region, VECSELs represent potential compact tunable laser sources, which are in high demand for various applications. In this region however, semiconductor structures are based on InP systems, such as ternary (e.g. InGaAs) or quaternary (e.g. InGaAsP) materials, which exhibit lower thermal conductivity than binary materials (e.g. GaAs, AlAs) grown on GaAs substrates for the 0.8 - 1 µm range. Moreover, because of the low refractive index contrast, a high number of InP/GaInAsP mirror pairs are required to form high reflectivity distributed Bragg reflectors. The resulting structures display high thermal impedance which prevents efficient heat removal from the active region to the heatsink. Therefore, the development of these structures has not been as successful as for the 0.8-1 µm region. However, room temperature (RT) operation of monolithically grown structures in simple compact two-mirror cavities has already been reported [7,8].

In this letter, we report the first application of a 1.55 µm RT-CW OP-VECSEL to high resolution spectroscopy, with the implementation of a sensitive ICLAS set-up analyzed by time-resolved Fourier transform spectroscopy.

The broadband laser consists of a ½-VCSEL structure designed for optical pumping at 980 nm by a commercial diode laser. The structure is composed of a Bragg mirror and a multiple-quantum-wells active region. The ½-VCSEL is mounted in an external cavity with standard dielectric mirrors in order to constitute the VECSEL device. The monolithic structure has been grown [7] on a 400 µm InP substrate by metal-organic vapor-phase epitaxy. The Bragg mirror of the ½-VCSEL structure is formed by 48 pairs of InP/InGaAsP quarterwave layers resulting in a reflectivity $R>99.8\%$. The active region is made of three half-wave groups of a strain-compensated package of two GaInAsP quantum wells. These layers are embedded between two

InP layers (0.75 and 2.25 λ) enhancing the heat dissipation, because of its good thermal conductivity ($K_C$ ~ 65 K.W$^{-1}$.m$^{-1}$). An anti-reflection coating has been deposited on the top surface of the ½-VCSEL to prevent a Fabry-Perot effect in the sub-cavity formed by the active region, thus to broaden the modal gain bandwidth in the external cavity [2].

As shown in Fig. 1, the ½-VCSEL is mounted onto a copper heat sink to control its temperature with a Peltier module, stabilized by a thermistor and a temperature controller. The structure is optically-pumped with a fiber-coupled 980 nm laser diode focused on a radius spot size of about 30 μm, matching the fundamental cavity mode. This size is fixed by thermal effects in the active region. A larger cavity mode would quadratically increase the pump power needed to reach the laser threshold, leading to higher temperatures inside the active region – as the thermal impedance decreases approximately linearly with pump radius - and preventing the occurrence of the laser oscillation. The L-shaped external cavity is formed by the ½-VCSEL, a concave (125 mm radius of curvature) and a plane dielectric high reflectivity ($R$>99.97%) wedged mirrors. The cavity has a 1-m long collimated arm, providing a longitudinal mode spacing of ~140 MHz. Laser oscillation in CW operation occurs around 1550 nm at 295 K with a threshold pump power of about 100 mW. The central wavelength can be tuned between 1545 and 1555 nm, through temperature variation of the ½-VCSEL and by moving the chip, thereby taking advantage of the growth rate non-uniformity. For ICLAS experiments, the pump current is modulated by a function generator in order to operate in quasi-CW regime. The inset in Fig. 1 shows typical shapes of the pump and laser pulses. The pump pulse duration is chosen between 100 μs and 1 ms, and the pump power is about 1.1 times the threshold. The copper heatsink temperature is maintained at 10°C. An absorption cell with Brewster-angled windows may be inserted in the collimated arm and fills about 60 % of the cavity. The light leaking outside the plane high reflectivity mirror (T<0.03%) is sent into a homemade high-resolution stepping-mode time-resolved Fourier-transform interferometer equipped with an infrasil beamsplitter and two InGaAs photodetectors. At a given path difference step, several time samples are recorded from a given laser pulse and several pulses are co-added. At the end of the experiment, there are as many interferograms as there are time samples [9]. Each time-component interferogram corresponds to a spectrum at a particular absorption path length.

Spectra have been recorded under various experimental conditions, with or without the absorption cell inside the laser cavity. Figure 2 exhibits 54 time-components from a 64-component time-resolved spectrum, separated from each other by 3.2 μs. The maximum generation time is 220 μs and the unapodized spectral resolution is 5.5 pm (22 10$^{-3}$ cm$^{-1}$). The width (FWHM) of the VECSEL emission spectrum is at most 7 nm for $t_g$ =50 μs and decreases [1] proportionally to $1/\sqrt{t_g}$ down to 3 nm for a generation time of 200 μs. The spectral width follows this theoretical square-root dependence at least up to 800 μs. Due to the temperature increase in the active region with the generation time, which is related to the thermal response of the InP-based structure, the central wavelength shifts towards longer wavelengths during a pump pulse. It varies from about 1545 nm up to 1549 nm at the end of a 200 μs pulse. The band structure on the first time-component spectrum presented in Fig.2 is the $P$ branch of the 30011-00001 band of $CO_2$, filling the cell at a pressure of about 200 Pa. Even though this spectral area is an atmospheric window, water absorption profiles are easily observed when the cell is removed from the cavity only filled then with ambient air. It is illustrated in the inset in Fig.2 representing a small portion of a time-component (path length: 65 km) belonging to another time-resolved spectrum. Figure 3 provides an expanded high-resolution portion of the first time-component spectrum of Fig 2 . The $P(28)e$ to $P(44)e$ lines of the 30011-00001 band of $CO_2$ are observed at 9 km path length (taking into account the 60% cavity filling ratio).

In the spectra presented in Fig. 2, signal to peak-to-peak noise ratio is at best of the order of 300 and rms noise of the order of 0.1 %, on the higher time components spectra. The absorption path length is at most of 65 km. Consequently, the minimum detectable absorption coefficient is about $5 \ 10^{-10}$ cm$^{-1}$ which is consistent with the previous results for ICLAS with VECSEL sources [2-4]. As the experimental time is 2h10 min for about 95 000 effective independent spectro-temporal elements, the noise equivalent absorption coefficient at one second averaging is $1.5 \ 10^{-10}$ cm$^{-1}$.Hz$^{-1/2}$ per spectral element. This result compares well with highly efficient recent cavity ring down reports [11-13] in the same spectral region. The present experiment is fully automated and provides a global consistency of the measurements.

Future directions include developing a more compact set-up. Improved VECSEL structures would be a benefit for the experiment: in particular, a broader tunability would be of interest for fundamental and applied spectroscopy, with an increased number of potential species to be probed. A better management of thermal effects would lead to better sensitivity and versatility. The use of a intra-cavity crystalline heatspreader bonded to the gain chip surface as demonstrated at 1549 nm by [14] would however introduce a spectral filter, but removal of the InP substrate and bonding of ½ VCSEL structure on a substrate exhibiting a lower thermal impedance [5,15, 16] appears promising.

This work has been financially supported by an Action Concertée Nouvelles Méthodologies Analytiques et Capteurs. P. Delaye (LCFIO) and L. Morvan (Thalès) are warmly acknowledged for lending an infrared camera. N. Picqué's email address is nathalie.picque@ppm.u-psud.fr

Figure captions:

Figure 1: Experimental setup of the OP-VECSEL applied to ICLAS – Inset: Pump and laser pulses at 1.1 times the threshold.

Figure 2: Time-resolved spectrum of the dynamics of a 1.55 μm OP-VECSEL. The cell is filled with about 200 Pa of $CO_2$. Apodized spectral resolution is 20 pm (0.08 cm$^{-1}$) on this representation. Inset: Atmospheric water vapor lines at 65 km ($t_g$=220μs) absorption path length and 5.5 pm (22 10$^{-3}$ cm$^{-1}$) unapodized resolution extracted from another time-resolved spectrum with no cell in the cavity.

Figure 3: a) Positions and intensities of $CO_2$ absorption lines from HITRAN database [10]. b) Restricted portion of an experimental spectrum with 5.5 pm (22 10$^{-3}$ cm$^{-1}$) unapodized resolution and 9 km ($t_g$=50μs) absorption path length (with intracavity cell). c) Positions and intensities of $H_2O$ absorption lines from HITRAN database.

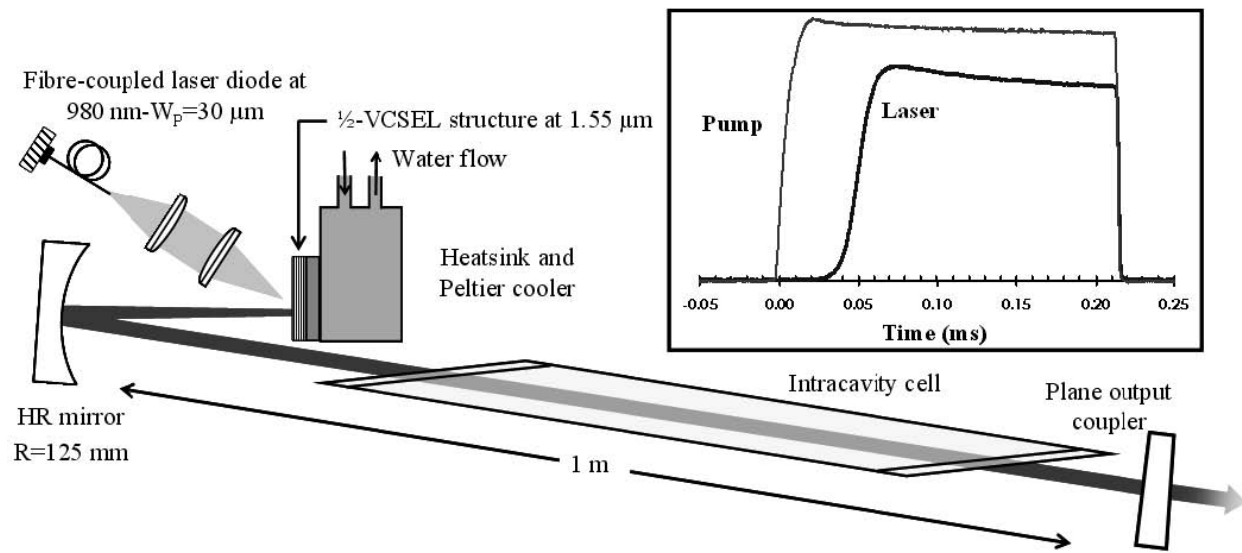

Figure 1: Experimental setup of the OP-VECSEL applied to ICLAS – Inset: Pump and laser pulses at 1.1 times the threshold.

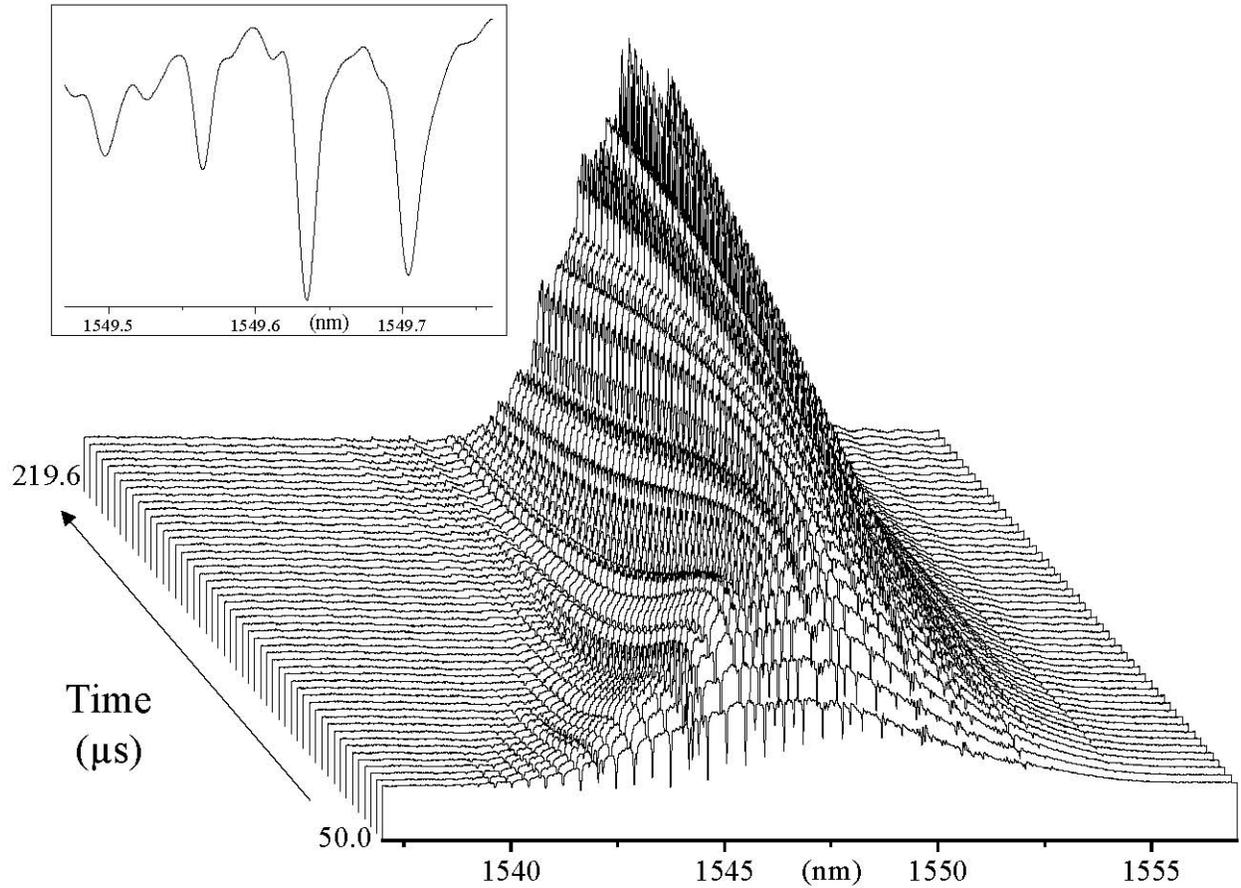

Figure 2: Time-resolved spectrum of the dynamics of a 1.55 μm OP-VECSEL. The cell is filled with about 200 Pa of $CO_2$. Apodized spectral resolution is 20 pm (0.08 cm$^{-1}$) on this representation. Inset: Atmospheric water vapor lines at 65 km ($t_g$ =220μs) absorption path length (without cell) and 5.5 pm (22 10$^{-3}$ cm$^{-1}$) unapodized resolution.

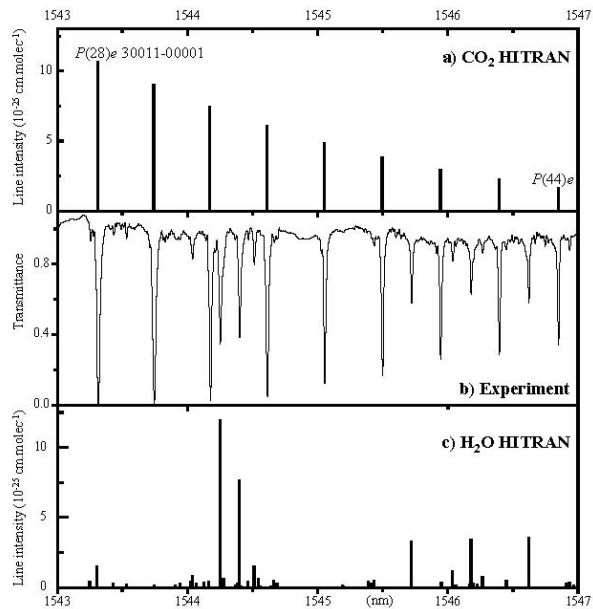

Figure 3: a) Positions and intensities of $CO_2$ absorption lines from HITRAN database [10]. b) Restricted portion of an experimental spectrum with 5.5 pm ($22\ 10^{-3}\ cm^{-1}$) unapodized resolution and 9 km ($t_g$ =50μs) absorption path length (with intracavity cell). c) Positions and intensities of $H_2O$ absorption lines from HITRAN database.